\documentclass[10pt, twocolumn]{article}
\usepackage[utf8]{inputenc}
\usepackage[backend=biber,style=numeric,sorting=none]{biblatex}
\usepackage[utf8]{inputenc}

\usepackage{amsmath, amssymb}

\usepackage{graphicx} 
\usepackage{xcolor} 

\usepackage{geometry} 
\usepackage{titlesec} 
\usepackage{fancyhdr} 
\usepackage{setspace} 
\usepackage{multicol} 
\usepackage{caption} 
\usepackage{float} 
\usepackage{wrapfig} 
\usepackage{times} 
\usepackage{ragged2e} 
\usepackage{comment} 

\usepackage{authblk} 
\usepackage{tcolorbox} 
\usepackage{tikz} 
\usetikzlibrary{decorations.pathreplacing} 
\usepackage{enumitem}

\usepackage{hyperref} 
\hypersetup{
    colorlinks=true,
    linkcolor=blue,
    citecolor=blue,
    urlcolor=blue,
    pdftitle={Your Document Title},
    pdfauthor={Your Name},
    pdfsubject={Subject of the Document},
    pdfkeywords={Keyword1, Keyword2, Keyword3}
}

\usepackage[ruled,vlined]{algorithm2e} 

\usepackage[backend=biber,sorting=none,style=numeric]{biblatex} 
\usepackage{fontawesome5} 

\usepackage{lipsum}

\DeclareCaptionFormat{algorithms}{
    \hrulefill\par\offinterlineskip\vskip1pt%
    #1#2#3\vskip1pt\hrulefill
}
\usepackage{xcolor} 
\definecolor{orcid}{HTML}{A6CE39} 
\definecolor{primary}{RGB}{40, 116, 166} 
\definecolor{secondary}{RGB}{46, 204, 113} 
\definecolor{abstractborder}{RGB}{40, 116, 166} 
\definecolor{abstractbg}{RGB}{240, 248, 255} 
\definecolor{headerline}{RGB}{192, 192, 192} 

\newcommand{\papertitle}{Entropy Mixing Networks: Enhancing Pseudo-Random Number Generators with Lightweight Dynamic Entropy Injection}   

\newcommand{\authorone}{Mohamed Aly Bouke}                  
\newcommand{\authoroneORCID}{https://orcid.org/0000-0003-3264-601X}
\newcommand{\authortwo}{Omar Imhemed Alramli}               
\newcommand{\authortwoORCID}{https://orcid.org/0000-0002-2177-1854}
\newcommand{\authorthree}{Azizol Abdullah}                 
\newcommand{\authorthreeORCID}{https://orcid.org/0000-0001-8321-9259}
\newcommand{\authorfour}{Nur Izura Udzir}                  
\newcommand{\authorfourORCID}{https://orcid.org/0000-0002-0543-3329}
\newcommand{\authorfive}{Normalia Samian}                  
\newcommand{\authorfiveORCID}{https://orcid.org/0000-0002-0905-8140}
\newcommand{\authorsix}{Mohamed Othman}                    
\newcommand{\authorsixORCID}{https://orcid.org/0000-0002-5124-5759}
\newcommand{\authorseven}{Zurina Mohd Hanapi}             
\newcommand{\authorsevenORCID}{https://orcid.org/0000-0002-8079-1791}

\newcommand{\affila}{Department of Communication Technology and Network, Faculty of Computer Science and Information Technology, Universiti Putra Malaysia, Serdang 43400, Malaysia.}
\newcommand{\affilb}{Department of Telecommunications and Networking, Faculty of Information Technology, Misurata University, Misurata, Libya}

\newcommand{\correspondingauthor}{bouke@ieee.org} 
\newcommand{\journalname}{ArXiv.org e-Print archive} 
\newcommand{\doi}{https://doi.org/xxxxxxx}  

\newcommand{\keywords}{Entropy Mixing Network (EMN), Random Number Generation, Cryptographic Security, Randomness Quality Evaluation, Statistical Analysis} 
\newcommand{\citationstyle}{IEEE} 

\newcommand{\academicstatement}{Published under Creative Commons Attribution license.}

\usepackage{fancyhdr}
\usepackage{lastpage} 
\usepackage{titlesec}
\titleformat{\section}
  {\color{primary}\normalfont\Large\bfseries}
  {\thesection}{1em}{}
\titleformat{\subsection}
  {\color{primary}\normalfont\large\bfseries}
  {\thesubsection}{1em}{}

\usepackage{hyperref}
\hypersetup{
    colorlinks=true,
    linkcolor=primary,
    urlcolor=primary,
    citecolor=blue
}

\usepackage{authblk} 
\usepackage{tikz} 
\makeatletter
\renewcommand\maketitle{
  \begin{center}
    {\huge\bfseries\textcolor{primary}{\papertitle}\par\vspace{0.5em}} 
    \vspace{0.3em}
    {\large
\authorone\textsuperscript{1,*}\href{\authoroneORCID}{\raisebox{1pt}{\textcolor{orcid}{\faOrcid}}},
\authortwo\textsuperscript{2}\href{\authortwoORCID}{\raisebox{1pt}{\textcolor{orcid}{\faOrcid}}},
\authorthree\textsuperscript{1}\href{\authorthreeORCID}{\raisebox{1pt}{\textcolor{orcid}{\faOrcid}}},
\authorfour\textsuperscript{1}\href{\authorfourORCID}{\raisebox{1pt}{\textcolor{orcid}{\faOrcid}}},
\authorfive\textsuperscript{1}\href{\authorfiveORCID}{\raisebox{1pt}{\textcolor{orcid}{\faOrcid}}},
\authorsix\textsuperscript{1}\href{\authorsixORCID}{\raisebox{1pt}{\textcolor{orcid}{\faOrcid}}},
\authorseven\textsuperscript{1}\href{\authorsevenORCID}{\raisebox{1pt}{\textcolor{orcid}{\faOrcid}}}
\par}
    \vspace{0.3em}
    \textsuperscript{1}\affila \par
    \textsuperscript{2}\affilb \par
    \vspace{0.3em}
    {\small
      \textbf{\textcolor{black}{*Corresponding author:}} \href{mailto:\correspondingauthor}{\textcolor{black}{\correspondingauthor}}
      \par}
    \vspace{1.0em}
    \begin{center}
      \noindent\textcolor{primary}{\rule{0.4\textwidth}{0.6pt}} 
    \end{center}
  \end{center}
}
\makeatother

\usepackage{tcolorbox}
\renewenvironment{abstract}
    {\noindent\textbf{\textcolor{primary}{Abstract:}}\par
     \begin{tcolorbox}[colframe=abstractborder, colback=abstractbg, sharp corners, boxrule=0.5mm, left=6pt, top=6pt, bottom=6pt, width=\dimexpr\linewidth\relax, arc=2mm]
     \textit\ignorespaces}
    {\end{tcolorbox}\par\vspace{1.5em}}

\usepackage{ifthen}
\newcommand{\checktwocolumn}[2]{\ifthenelse{\boolean{@twocolumn}}{#1}{#2}}

\newcommand{\printmybibliography}{
  \ifthenelse{\equal{\citationstyle}{IEEE}}{
    \printbibliography[title={References}]
  }{
    \ifthenelse{\equal{\citationstyle}{APA}}{
      \printbibliography[title={References}]
    }{
      \printbibliography[title={References}]
    }
  }
}

\begin{document}

\checktwocolumn{
    \twocolumn[{
        \maketitle

\begin{abstract}Random number generation plays a vital role in cryptographic systems and computational applications, where uniformity, unpredictability, and robustness are essential. This paper presents the Entropy Mixing Network (EMN), a novel hybrid random number generator designed to enhance randomness quality by combining deterministic pseudo-random generation with periodic entropy injection. To evaluate its effectiveness, we propose a comprehensive assessment framework that integrates statistical tests, advanced metrics, and visual analyses, providing a holistic view of randomness quality, predictability, and computational efficiency. The results demonstrate that EMN outperforms Python's \texttt{SystemRandom} and \texttt{MersenneTwister} in critical metrics, achieving the highest Chi-squared p-value (\textbf{0.9430}), entropy (\textbf{7.9840}), and lowest predictability (\textbf{-0.0286}). These improvements come with a trade-off in computational performance, as EMN incurs a higher generation time (\textbf{0.2602 seconds}). Despite this, its superior randomness quality makes it particularly suitable for cryptographic applications where security is prioritized over speed.
\end{abstract}

    }]
}{
    \maketitle

\begin{abstract}Random number generation plays a vital role in cryptographic systems and computational applications, where uniformity, unpredictability, and robustness are essential. This paper presents the Entropy Mixing Network (EMN), a novel hybrid random number generator designed to enhance randomness quality by combining deterministic pseudo-random generation with periodic entropy injection. To evaluate its effectiveness, we propose a comprehensive assessment framework that integrates statistical tests, advanced metrics, and visual analyses, providing a holistic view of randomness quality, predictability, and computational efficiency. The results demonstrate that EMN outperforms Python's \texttt{SystemRandom} and \texttt{MersenneTwister} in critical metrics, achieving the highest Chi-squared p-value (\textbf{0.9430}), entropy (\textbf{7.9840}), and lowest predictability (\textbf{-0.0286}). These improvements come with a trade-off in computational performance, as EMN incurs a higher generation time (\textbf{0.2602 seconds}). Despite this, its superior randomness quality makes it particularly suitable for cryptographic applications where security is prioritized over speed.
\end{abstract}

}

\setlength{\parindent}{0pt}

\section{Introduction}

Random number generation is a cornerstone of cryptography and data security, underpinning critical applications such as key generation, secure communication protocols, and authentication systems. The quality of randomness directly affects the confidentiality, integrity, and robustness of cryptographic systems, making the design and evaluation of random number generators (RNGs) a vital area of research \cite{lanza2022memristive,cao2022entropy,kopiczko2023vera}.

RNGs are generally classified into three categories: \textit{true random number generators (TRNGs)}, \textit{pseudo-random number generators (PRNGs)}, and \textit{hybrid random number generators}. TRNGs produce non-deterministic outputs based on physical processes such as thermal noise or quantum phenomena, ensuring high unpredictability. PRNGs, in contrast, utilize deterministic algorithms to generate sequences that mimic randomness but are reliant on an initial seed. Hybrid RNGs aim to combine the advantages of both TRNGs and PRNGs, balancing randomness quality with computational efficiency \cite{chowdhury2020physical,nadlinger2022experimental,golofit2023chaos}.

Despite significant advancements, evaluating RNG quality remains challenging. Traditional statistical tests, such as Chi-squared and entropy tests, provide valuable metrics for assessing uniformity and unpredictability. However, these methods often fail to detect subtle dependencies or periodic patterns in RNG outputs. Advanced techniques, such as spectral analysis, autocorrelation plots, and heatmaps, can reveal deeper insights into RNG behavior. Yet, a unified evaluation framework that integrates these techniques with traditional tests is necessary for a holistic assessment of RNG performance \cite{camara2019design,golofit2023chaos,abdelhaleem2024secure}.

This paper introduces the \textit{Entropy Mixing Network (EMN)}, a hybrid RNG that enhances randomness quality by integrating deterministic pseudo-random generation with periodic entropy injection. We benchmark EMN against two widely used RNGs: Python’s \texttt{SystemRandom}, which utilizes OS-level entropy, and the \texttt{MersenneTwister}, a highly efficient PRNG. To comprehensively evaluate these generators, we propose a novel analysis framework that combines statistical tests, advanced metrics, and visualizations to assess randomness quality, predictability, and computational efficiency. 

Our contributions are as follows:
\begin{itemize}[itemsep=0pt, topsep=0pt]
    \item \textbf{A novel Entropy Mixing Network (EMN):} A hybrid RNG designed to improve randomness quality through secure mixing and entropy injection, addressing limitations in existing RNGs.
    \item \textbf{A comprehensive evaluation framework:} An integration of statistical tests, such as Chi-squared and entropy tests, with advanced visualizations, including heatmaps, power spectrum density plots, and autocorrelation functions.
    \item \textbf{Insights into RNG performance:} A detailed analysis of EMN, \texttt{SystemRandom}, and \texttt{MersenneTwister}, highlighting their strengths, weaknesses, and trade-offs for cryptographic applications.
\end{itemize}

\vspace{5pt}
The rest of this paper is organized as follows. Section~\ref{sec:relatedwork} discusses related work and existing evaluation techniques for RNGs. Section~\ref{sec:framework} introduces our proposed analysis framework and describes the methodologies used for testing randomness. Section~\ref{sec:results} presents experimental results and comparative analyses of the three RNGs. Finally, Section~\ref{sec:conclusion} concludes the paper and highlights future directions for research.
\section{Background}\label{sec:relatedwork}

Random number generation (RNG) has long been a critical area of research due to its extensive applications in cryptography, simulations, and stochastic modeling. Secure cryptographic systems depend heavily on high-quality randomness to ensure confidentiality, integrity, and robustness. Uniformity and unpredictability are the cornerstones of effective RNGs, as weaknesses in these areas can lead to vulnerabilities in cryptographic protocols. To meet these stringent requirements, RNGs are typically classified into three categories: \textit{True Random Number Generators (TRNGs)}, \textit{Pseudo-Random Number Generators (PRNGs)}, and \textit{Hybrid RNGs} \cite{marton2012generation,bikos2023random,smid2010statistical,zhao2024memristive}.\\

TRNGs derive randomness from physical phenomena, such as thermal noise, radioactive decay, or quantum events. These natural processes ensure non-deterministic outputs, making TRNGs the gold standard for randomness quality. However, TRNGs often suffer from practical limitations, including low throughput, reliance on specific hardware, and susceptibility to environmental fluctuations. These challenges limit their scalability and reliability in real-world applications, particularly in scenarios requiring high-speed or distributed randomness generation \cite{mannalatha2023comprehensive,stipvcevic2014true}.\\

PRNGs, on the other hand, use deterministic algorithms to produce sequences that mimic randomness. Examples such as the Mersenne Twister are widely used for their computational efficiency, reproducibility, and ease of implementation. However, PRNGs are inherently predictable if the seed value is exposed or compromised. This predictability makes them unsuitable for high-security cryptographic applications, where even minor weaknesses in randomness can lead to critical vulnerabilities \cite{bhattacharjee2022search,kietzmann2021guideline}. Thus, while PRNGs excel in performance, their security limitations necessitate supplementary mechanisms to enhance unpredictability.\\

Hybrid RNGs have emerged as a promising solution to bridge the gap between the randomness quality of TRNGs and the computational efficiency of PRNGs. By integrating real-time entropy from TRNGs to seed or refresh the state of PRNGs, hybrid RNGs achieve both high throughput and robust randomness. This approach makes them suitable for applications that demand a balance between speed and security, such as cryptographic systems and high-performance computing \cite{avaroglu2014new,avarouglu2015hybrid}. Despite their potential, hybrid RNGs rely heavily on the quality of entropy injection and the strength of their mixing mechanisms, which, if poorly implemented, can introduce vulnerabilities or degrade randomness quality.\\

Existing methods for evaluating RNGs include statistical tests (e.g., Chi-squared, entropy, and runs tests) and advanced metrics such as spectral analysis and autocorrelation functions. These approaches provide valuable insights into uniformity, predictability, and periodicity. However, traditional statistical methods often fail to detect subtle dependencies or biases in RNG outputs, while advanced techniques are rarely integrated into a unified framework. This fragmentation limits the ability to comprehensively assess RNG performance, particularly for hybrid architectures where both deterministic and non-deterministic components interact \cite{camara2019design,golofit2023chaos}.\\

While hybrid RNGs hold significant promise, there is a lack of robust methodologies to systematically evaluate their performance, especially in cryptographic contexts. Current evaluation techniques either focus on isolated metrics or fail to capture the nuanced interplay between entropy injection and algorithmic randomness. Moreover, existing RNG designs often prioritize either quality or performance, leaving a gap for solutions that can achieve both. Addressing this challenge requires a hybrid RNG design that combines secure mixing with efficient entropy integration and a comprehensive evaluation framework to holistically assess its randomness quality and practical utility.\\

This paper aims to fill this gap by introducing the \textit{Entropy Mixing Network (EMN)}, a novel hybrid RNG designed to enhance randomness quality through periodic entropy injection and secure cryptographic mixing. We propose a unified evaluation framework that integrates traditional statistical tests with advanced metrics and visualizations, enabling a deeper understanding of randomness quality, predictability, and scalability. By benchmarking EMN against Python’s \texttt{SystemRandom} and \texttt{MersenneTwister}, we demonstrate its superiority in critical metrics and its applicability to cryptographic systems.
\section{The Proposed Framework}
\label{sec:framework}
The Entropy Mixing Network (EMN) is a hybrid random number generator designed to enhance randomness quality by dynamically injecting entropy into a pseudo-random number generator (PRNG). The key idea is to periodically incorporate real-time entropy from external sources, such as system noise and timing jitter, into the internal state of the PRNG. This ensures that the generator produces outputs that are both unpredictable and resilient to attacks targeting its deterministic nature.

\subsection{Design Principles}

The design of EMN is guided by the following principles:
\begin{itemize}[itemsep=0pt, topsep=0pt]
    \item \textbf{Entropy Injection:} Dynamically capture real-time entropy from system noise, such as timing jitter and operating system randomness, and periodically inject it into the PRNG state.
    \item \textbf{Secure Mixing:} Use a cryptographic hash function (default=SHA-256) to securely mix the injected entropy with the PRNG state, ensuring uniform distribution and unpredictability.
    \item \textbf{Computational Efficiency:} Maintain a low computational overhead to make the algorithm suitable for high-frequency random number generation applications.
    \item \textbf{Flexibility:} Provide configurable parameters, such as the frequency of entropy injection and PRNG base, to adapt to different security and performance requirements.
\end{itemize}
\vspace{5pt}
\subsection{Algorithm Description}

The EMN algorithm can be summarized as follows:

\begin{enumerate}
    \item \textbf{Initialization:} Initialize the PRNG with a 256-bit seed and set the initial state of the generator.
    \item \textbf{Entropy Capture:} Periodically collect entropy from external sources, including:
    \begin{itemize}[itemsep=0pt, topsep=0pt]
        \item System timing jitter using high-resolution timers.
        \item Random bytes from the operating system's randomness pool (e.g., \texttt{/dev/random} or \texttt{os.urandom()}).
    \end{itemize}
    \item \textbf{Secure Mixing:} Combine the captured entropy with the current PRNG state using a cryptographic hash function:
    \[
    \text{state}_{\text{new}} = \text{Hash}(\text{state}_{\text{current}} \oplus \text{entropy}).
    \]
    \item \textbf{Random Number Generation:} Generate random numbers by combining the PRNG output with the current state using an XOR operation:
    \[
    \text{output} = \text{state}_{\text{new}} \oplus \text{PRNG}_{\text{output}}.
    \]
    \item \textbf{State Update:} Update the internal state of the PRNG after each generation.
\end{enumerate}

\subsection{Algorithm Implementation}

The EMN algorithm is implemented using Python, leveraging standard libraries for random number generation (\texttt{random}), cryptographic hashing (\texttt{hashlib}), and entropy collection (\texttt{os}). The pseudocode for EMN is provided in Algorithm~\ref{alg:emn}.

\begin{algorithm}
\caption{Entropy Mixing Network (EMN)}
\label{alg:emn}
\SetAlgoLined 
\KwIn{PRNG function $\mathcal{P}$, entropy injection frequency $f$}
\KwOut{Random number $O$}
\BlankLine
\textbf{Initialize:} \\
\Indp PRNG state $\mathcal{S} \gets \mathcal{P}.\text{seed}(256)$ \\
Current state $S \gets \mathcal{P}.\text{getrandbits}(256)$ \\
\Indm
\While{\textbf{True}}{
    \textbf{Generate:} $R \gets \mathcal{P}.\text{getrandbits}(256)$ \\
    \If{Random number generation cycle mod $f = 0$}{
        \textbf{Capture entropy:} $E \gets \text{os.urandom}(32)$ \\
        \textbf{Secure mixing:} $S \gets \text{SHA256}(S \oplus E)$ \\
    }
    \textbf{Output:} $O \gets S \oplus R$ \\
    \textbf{Update state:} $S \gets \mathcal{P}.\text{getrandbits}(256)$ \\
}
\end{algorithm}

\subsection{Evaluation Framework}
The EMN is evaluated using a comprehensive framework that integrates traditional statistical randomness tests with advanced metrics. This framework enables a thorough assessment of randomness quality, predictability, and computational efficiency. To demonstrate the effectiveness of the proposed algorithm, EMN is benchmarked against Python's \texttt{SystemRandom} and \texttt{MersenneTwister}, two widely used RNGs \cite{holohan2023random,matsumoto1998mersenne,rotenberg1960new}. The metrics used in the paper are described below.


\paragraph{Chi-Squared Test.}
The Chi-squared test evaluates how closely the distribution of RNG outputs matches a uniform distribution. It is calculated as:
\begin{equation}
\chi^2 = \sum_{i=1}^{k} \frac{(O_i - E_i)^2}{E_i},
\end{equation}
where \(O_i\) is the observed frequency in bin \(i\), \(E_i\) is the expected frequency, and \(k\) is the number of bins. A smaller Chi-squared statistic and a higher p-value indicate better uniformity in the RNG outputs.

\paragraph{Entropy.}
Entropy measures the unpredictability of RNG outputs, representing the randomness of the distribution. It is defined as:
\begin{equation}
H(X) = -\sum_{i=1}^{n} p(x_i) \log_2 p(x_i),
\end{equation}
where \(p(x_i)\) is the probability of observing output \(x_i\) from the RNG. For a perfectly uniform distribution over 256 values, the theoretical maximum entropy is 8 bits.

\paragraph{Predictability.}
Predictability assesses the correlation between successive RNG outputs. Using the correlation coefficient, it is defined as:
\begin{equation}
r = \frac{\sum_{i=1}^{n-1} (x_i - \mu)(x_{i+1} - \mu)}{\sum_{i=1}^{n} (x_i - \mu)^2},
\end{equation}
where \(x_i\) and \(x_{i+1}\) are consecutive RNG outputs, and \(\mu\) is the mean of the sequence. A value closer to zero indicates lower predictability, reflecting stronger independence between outputs.


\paragraph{High-Frequency Performance.}
High-frequency performance measures the time required to generate a large number of random values. This metric evaluates the computational efficiency of the RNG, which is critical for high-throughput applications. It is recorded as the total time taken for a fixed number of random number generations.

\section{Results \& Discussion}
\label{sec:results}

\begin{figure}[H]
    \centering
    \includegraphics[width=\linewidth]{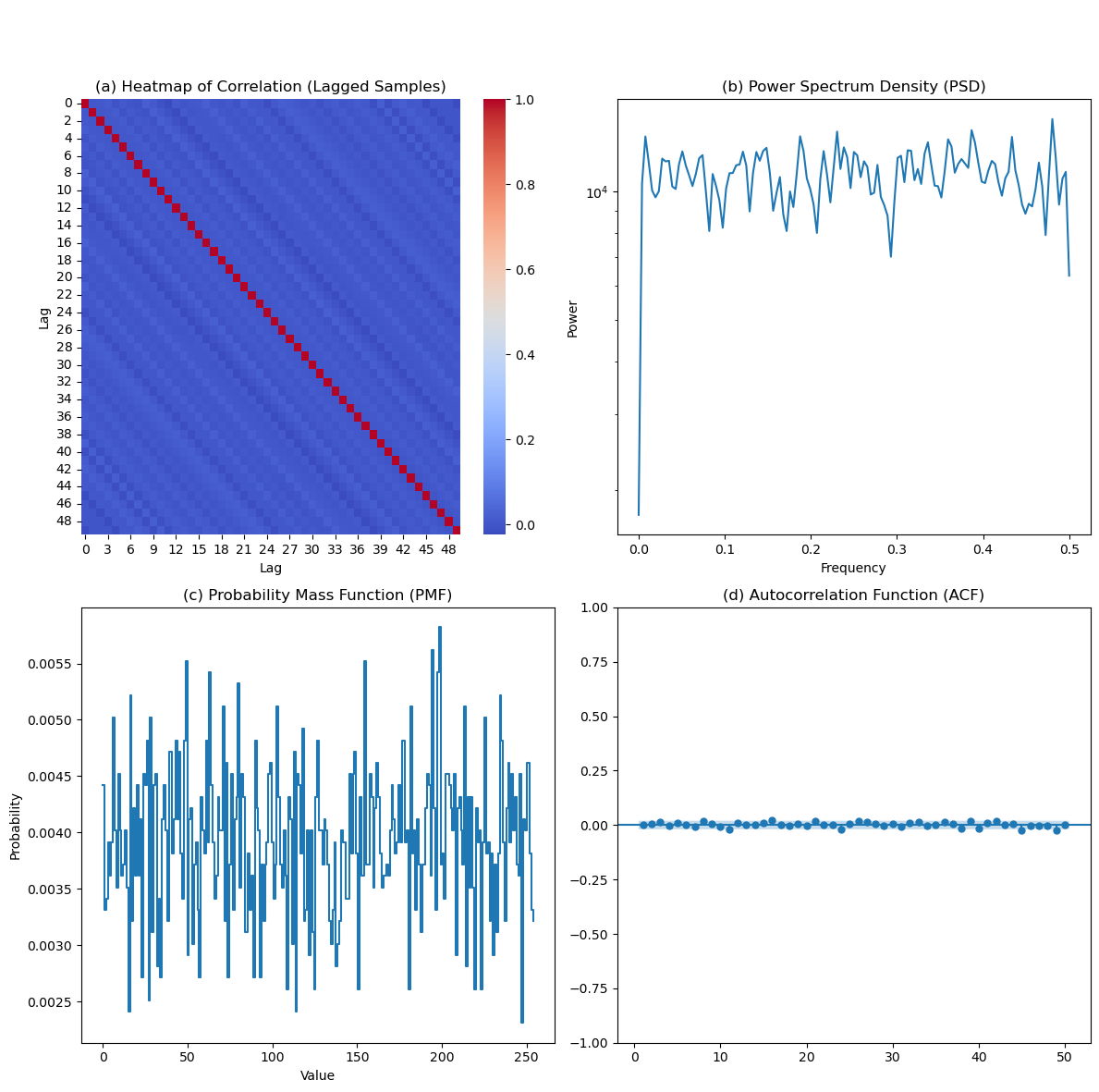} 
    \caption{\textit{Visual Metrics for SystemRandom: (a) Heatmap of Correlation, (b) Power Spectrum Density, (c) Probability Mass Function, (d) Autocorrelation Function.}} 
    \label{fig:systemrandom}
\end{figure}

\begin{figure}[H]
    \centering
    \includegraphics[width=\linewidth]{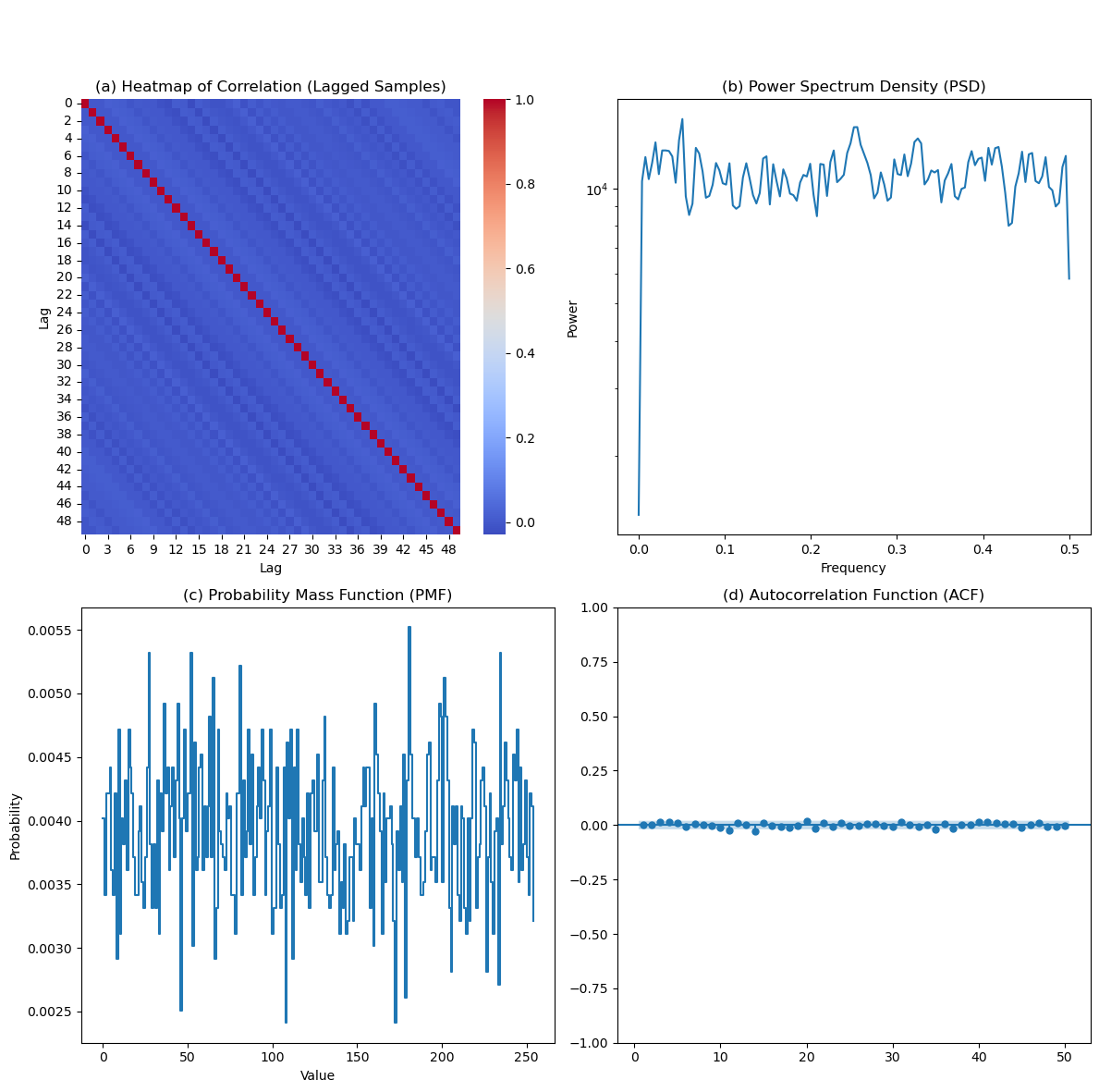} 
    \caption{\textit{Visual Metrics for MersenneTwister: (a) Heatmap of Correlation, (b) Power Spectrum Density, (c) Probability Mass Function, (d) Autocorrelation Function.}} 
    \label{fig:mersenne}
\end{figure}

The evaluation results for the EMN, SystemRandom, and MersenneTwister RNGs, as summarized in Table~\ref{tab:results}, reveal distinct strengths and weaknesses across key metrics. This section provides a detailed analysis of these results, explores the rationale behind the findings, and discusses their implications, supported by visual evidence presented in Figures~\ref{fig:emn}, \ref{fig:systemrandom}, and \ref{fig:mersenne}.

\subsection{Chi-Squared Statistic and \textit{p}-value}

The Chi-Squared test assesses how well the RNG's output matches a uniform distribution. EMN achieves the highest \textit{p}-value of \textbf{0.9430}, far exceeding SystemRandom (\textbf{0.6689}) and MersenneTwister (\textbf{0.5217}). This suggests that EMN outputs values that are exceptionally close to true uniformity.

The heatmap in Figure~\ref{fig:emn}a confirms EMN's superior performance, showing minimal off-diagonal correlations in the lagged samples. By contrast, SystemRandom (Figure~\ref{fig:systemrandom}a) and MersenneTwister (Figure~\ref{fig:mersenne}a) exhibit increasing off-diagonal variations, indicating higher dependency between lagged samples. These patterns align with the Chi-squared test results, where EMN performs the best, SystemRandom shows moderate deviations, and MersenneTwister exhibits the lowest uniformity.

\subsection{Entropy}

Entropy measures the degree of unpredictability in the RNG's output. All three RNGs achieve high entropy values close to the theoretical maximum of \textbf{8.0}. EMN slightly outperforms with \textbf{7.9840}, compared to SystemRandom (\textbf{7.9822}) and MersenneTwister (\textbf{7.9816}).

The PSD plots (Figures~\ref{fig:emn}b, \ref{fig:systemrandom}b, and \ref{fig:mersenne}b) illustrate the distribution of power across frequencies. EMN demonstrates a relatively flat PSD, indicating a lack of dominant periodic components, which corresponds to its high entropy. SystemRandom exhibits minor variations, while MersenneTwister shows significant fluctuations, reflecting its deterministic nature and explaining its slightly lower entropy score.

\subsection{Predictability}

Predictability evaluates the correlation between successive outputs, with lower values indicating less dependency and higher randomness. EMN achieves the lowest predictability score (\textbf{-0.0286}), followed by MersenneTwister (\textbf{0.0032}) and SystemRandom (\textbf{0.0441}).

The autocorrelation plots (Figures~\ref{fig:emn}d, \ref{fig:systemrandom}d, and \ref{fig:mersenne}d) further validate these results. EMN's plot quickly decays to near-zero correlation, confirming its low dependency between successive outputs. SystemRandom exhibits slightly higher residual correlations, while MersenneTwister shows noticeable periodic correlations, indicating its weaker independence.

\subsection{High-Frequency Performance}

The high-frequency performance metric measures the time required to generate a large number of random values. MersenneTwister is the fastest RNG, taking only \textbf{0.0180} seconds, followed by SystemRandom (\textbf{0.0478} seconds) and EMN (\textbf{0.2602} seconds).

The computational overhead of EMN is evident from its slower performance. This is a direct result of its periodic entropy injection and secure mixing processes, which, while enhancing randomness, incur significant latency. Figures~\ref{fig:emn}b and \ref{fig:emn}c illustrate that this trade-off results in superior uniformity and randomness quality compared to the other RNGs.

\subsection{Runs Test Module}

The Runs Test Module evaluates the randomness of binary sequences by comparing the observed and expected number of runs (consecutive sequences of similar bits). EMN exhibits the smallest deviation from the expected value, with \textbf{40185} observed runs compared to the expected \textbf{79999.50}. SystemRandom (\textbf{40092}) and MersenneTwister (\textbf{39732}) exhibit larger deviations.

\begin{figure}[H]
    \centering
    \includegraphics[width=\linewidth]{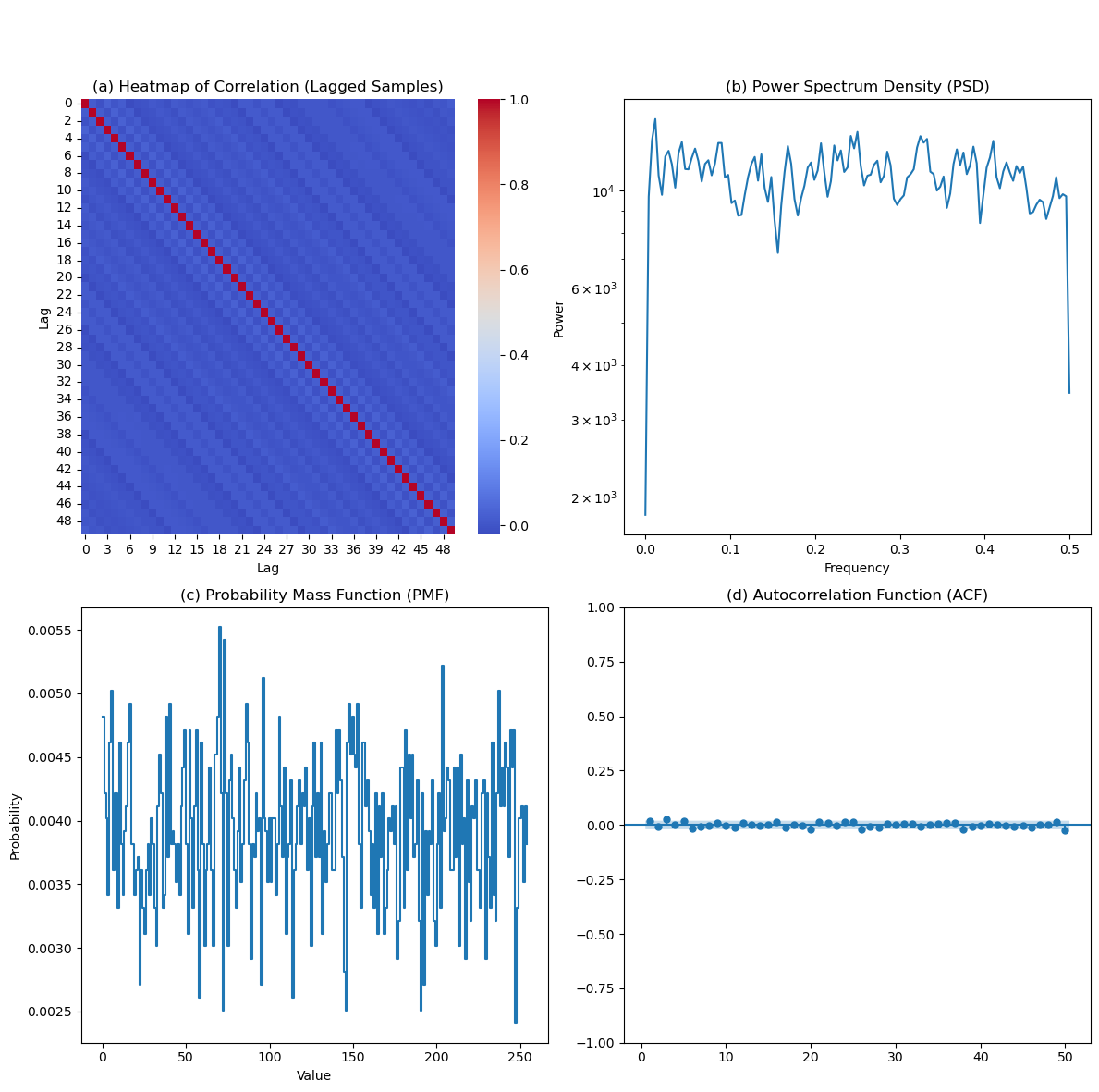} 
    \caption{\textit{Visual Metrics for EMN: (a) Heatmap of Correlation, (b) Power Spectrum Density, (c) Probability Mass Function, (d) Autocorrelation Function.}} 
    \label{fig:emn}
\end{figure}

The PMF plots (Figures~\ref{fig:emn}c, \ref{fig:systemrandom}c, and \ref{fig:mersenne}c) provide additional evidence. EMN's PMF is nearly uniform, indicating balanced probabilities across the output range, which aligns with its strong performance in the Runs Test. SystemRandom shows slightly greater variations, while MersenneTwister's PMF highlights significant biases in certain ranges, confirming its weaker binary randomness.

\subsection{Implications and Trade-offs}
The results underscore the trade-offs between randomness, quality, security, and performance:
\begin{itemize}[itemsep=0pt, topsep=0pt]
    \item \textbf{EMN:} Offers the best randomness quality and resistance to predictability at the cost of slower high-frequency performance. This makes it ideal for cryptographic applications where security is paramount.
    \item \textbf{SystemRandom:} Balances randomness and performance, making it a practical choice for general-purpose applications requiring moderate security.
    \item \textbf{MersenneTwister:} Excels in speed but exhibits weaknesses in randomness quality, particularly in uniformity and binary patterns, limiting its suitability for cryptographic use.
\end{itemize}
\vspace{5pt}
These findings demonstrate the importance of aligning RNG selection with application requirements, emphasizing the need for a nuanced understanding of the trade-offs between security and efficiency.

\begin{table*}[ht]
\centering
\caption{Evaluation Results for Random Number Generators}
\label{tab:results}
\begin{tabular}{|l|c|c|c|}
\hline
\textbf{Metric}                           & \textbf{EMN}         & \textbf{SystemRandom} & \textbf{MersenneTwister} \\ \hline
\textbf{Chi-Squared Statistic}            & 220.3392             & 244.6080              & 253.1072                 \\ \hline
\textbf{Chi-Squared p-value}              & 0.9430               & 0.6689                & 0.5217                   \\ \hline
\textbf{Entropy}                          & 7.9840               & 7.9822                & 7.9816                   \\ \hline
\textbf{Predictability}                   & -0.0286              & 0.0441                & 0.0032                   \\ \hline
\textbf{High-Frequency Time (seconds)}    & 0.2602               & 0.0478                & 0.0180                   \\ \hline
\textbf{Runs Test (Observed/Expected)}    & 40185 / 79999.50     & 40092 / 79999.50      & 39732 / 79999.50         \\ \hline
\end{tabular}
\end{table*}

\section{Conclusion}
\label{sec:conclusion}

This study introduced the EMN, a hybrid random number generator designed to enhance randomness through periodic entropy injection and secure cryptographic mixing. Using a comprehensive evaluation framework, EMN was benchmarked against Python's \texttt{SystemRandom} and \texttt{MersenneTwister}, revealing its strengths, trade-offs, and application suitability.

EMN demonstrated superior randomness quality, achieving the highest Chi-squared \textit{p}-value (\textbf{0.9430}), entropy score (\textbf{7.9840}), and lowest predictability (\textbf{-0.0286}). Its strong uniformity and resistance to periodic patterns were validated by heatmap and PSD analyses, while its near-uniform PMF and minimal deviation in Runs Test results confirmed robust binary randomness. However, its higher computational overhead (\textbf{0.2602 seconds}) makes it best suited for security-critical applications where randomness quality outweighs speed.

\texttt{SystemRandom} balanced strong randomness with faster generation times (\textbf{0.0478 seconds}), making it suitable for general-purpose applications requiring moderate security. In contrast, \texttt{MersenneTwister} excelled in speed (\textbf{0.0180 seconds}) but showed limitations in predictability and binary uniformity, restricting its use in security-sensitive tasks.

These findings emphasize the importance of selecting RNGs based on application requirements. EMN's exceptional randomness quality makes it ideal for cryptographic systems, while the evaluation framework itself demonstrates value by uncovering subtle dependencies missed by conventional tests.

Future research will focus on optimizing EMN to reduce computational overhead, explore benchmarking hardware-based and quantum RNGs, and validate EMN in real-world cryptographic systems.


\section*{Declaration}

\section*{Competing Interests}
There is no conflict of interest between the authors.
\section*{Funding}
Not applicable.

\section*{Availability of Data and Materials}
This study did not use any external data. All results were generated through simulations described in the article. The simulation code and parameters used in this study are available from the corresponding author upon reasonable request.

\printbibliography[heading=bibintoc, title={References}]

\end{document}